\documentclass[twocolumn,floatfix,showpacs,prl,aps,tightenlines
]{revtex4}
\usepackage{graphicx}
\usepackage{color}
\usepackage{psfrag}
\usepackage{dcolumn}
\usepackage{bm}

\newcommand{\bea}{\begin{eqnarray}}
\newcommand{\eea}{\end{eqnarray}}
\newcommand{\beq}{\begin{equation}}
\newcommand{\eeq}{\end{equation}}
\newcommand{\KMS}{\rm km\,s^{-1}}

\begin{document}

\title{Intermediate Mass Ratio Black Hole Binaries:\\
Numerical Relativity meets Perturbation Theory}

\author{Carlos O. Lousto}
\affiliation{Center for Computational Relativity and Gravitation,
School of Mathematical Sciences,
Rochester Institute of Technology, 85 Lomb Memorial Drive, Rochester,
 New York 14623}

\author{Hiroyuki Nakano}
\affiliation{Center for Computational Relativity and Gravitation,
School of Mathematical Sciences,
Rochester Institute of Technology, 85 Lomb Memorial Drive, Rochester,
 New York 14623}

\author{Yosef Zlochower} 
\affiliation{Center for Computational Relativity and Gravitation,
School of Mathematical Sciences,
Rochester Institute of Technology, 85 Lomb Memorial Drive, Rochester,
 New York 14623}

\author{Manuela Campanelli}
\affiliation{Center for Computational Relativity and Gravitation,
School of Mathematical Sciences,
Rochester Institute of Technology, 85 Lomb Memorial Drive, Rochester,
 New York 14623}

\date{\today}

\begin{abstract} 
We study black-hole binaries in the intermediate-mass-ratio regime
$0.01\lesssim q \lesssim 0.1$ with a new technique that makes use of
nonlinear numerical trajectories and efficient perturbative evolutions
to compute waveforms at large radii for the leading and nonleading
modes. As a proof-of-concept, we compute  waveforms for $q=1/10$.
We discuss applications of these techniques for LIGO/VIRGO
data analysis and the possibility that our technique can be extended
to produce accurate waveform templates from a modest number of
fully-nonlinear numerical simulations.
\end{abstract}

 \maketitle

\noindent
{\it Introduction:}
The dramatic breakthroughs in the numerical techniques to evolve
black-hole binaries (BHB)~\cite{Pretorius:2005gq, Campanelli:2005dd,
Baker:2005vv} transformed the field and made it possible to begin
constructing waveform templates for use in LIGO and VIRGO gravitational
wave data analysis and detection from highly-accurate, fully-nonlinear
simulations. This promises to greatly aid in the detection and analysis of
gravitational waves from astrophysical sources~\cite{Aylott:2009ya}.

While LISA's sensitivity band is well suited to observe supermassive
BHBs, ground based observatories are more sensitive to BHBs with
masses up to  a few hundred solar masses. Both of these types of
BHBs are most likely
to have mass ratios in the range 1:10 - 1:100~\cite{Volonteri:2008gj}.
This regime is  hard to model with fully-nonlinear numerical
simulations, which thus far have focused on the comparable mass
regime. On the other hand, BHBs in the very small mass ratio regime
can, in principle, be modeled accurately with perturbation theory, and
significant theoretical effort was
dedicated to the consistent computation of the self-force on a
point-like source orbiting around Schwarzschild and Kerr black holes
(see \cite{CQGVol22No15} for a review of new
developments in the field).
These perturbative techniques allow one to formally compute geodesic
deviations to second order. Although secular effects still need to be
quantified, these techniques are expected to provide a good approximation
for mass ratios $q=m_1/m_2<1/100$.

Fully-nonlinear BHB simulations with small mass-ratios are challenging and
the smallest mass ratios published so far are $q=1/10$ in the
nonspinning case~\cite{Gonzalez:2008bi, Baker:2008mj} and $q=1/8$ for
highly-spinning BHBs~\cite{Lousto:2008dn}.  Most fully-nonlinear
numerical simulations published have been in the comparable mass
regime, and consequently, most of the comparisons between NR and
post-Newtonian (PN) waveforms have been in this regime (for
comparisons of NR and PN waveforms from generic binaries, see
\cite{Campanelli:2008nk, Szilagyi:2009qz}  and references therein).
Similarly, phenomenological \cite{Ajith:2009bn} and EOBNR
\cite{Pan:2009wj} modeling of waveforms are based on comparable-mass
BHB simulations.  The intermediate region of $1/100<q<1/10$ provides a
unique opportunity to compare fully-nonlinear numerical and
perturbative techniques.  In this letter we introduce a novel approach
to deal with this intermediate-mass-ratio regime. We perform a `proof
of principle' comparison, to determine an upper bound on
the error in our technique, between fully-nonlinear numerical simulations
and perturbative ones in the borderline case of a nonspinning BHB
system with mass ratio $q=1/10$.

The key element introduced in our approach is the use of the
numerically generated trajectories (in the moving puncture
approach~\cite{Campanelli:2005dd, Baker:2005vv}) of a BHB during the
late inspiral regime in a perturbative calculation of the waveform
emitted by the binary.\\
\\
\noindent
{\it Full Numerical Simulations:}
In order to approximate the late orbital parameters of a quasicircular
inspiral we computed the momentum and puncture-position parameters
using a resummed (particle limit) 3.5 PN evolution of
a $q=1/10$, non-spinning binary from $r=50M$ to $r=7.25M$.  To compute
the numerical initial data, we use the puncture
approach~\cite{Brandt97b} along with the {\sc
TwoPunctures}~\cite{Ansorg:2004ds} thorn.  We chose the puncture mass
parameters such that the system had a total ADM mass $M = 1$ and mass ratio
$q=1/10$.

  We evolved this BHB data-set using the {\sc
LazEv}~\cite{Zlochower:2005bj} implementation of the moving puncture
approach~\cite{Campanelli:2005dd,Baker:2005vv}.  
Our code used the Cactus toolkit~\cite{Cactusweb} and the 
Carpet~\cite{Schnetter-etal-03b} mesh refinement driver to
provide a `moving boxes' style mesh refinement. We used a modified
gauge condition suggested in~\cite{Mueller:2009jx}. Our simulation used
11 levels of refinement (around the smaller components), with central
 resolutions as high as $M/368.64$, and 9 levels of refinement around
the larger component. The outer boundaries were located at $400M$ and
the resolution in the boundary zone was $h=2.7777M$ for our finest
resolution run. The BHB performs four orbits prior to merger, which
occurs roughly $400M$ after the start of the simulation (see
 Table~\ref{table:NID}).

\begin{table}
\caption{Initial puncture data, and final remnant parameters}
\begin{ruledtabular}
\begin{tabular}{lclc}
$x_1$ & 6.60438 & $x_2$ & -0.671518 \\
$m_1^p$ & $8.43895\times10^{-2}$ & $m_2^p$ & $0.907039$ \\
$p_x$ & $-3.2671\times10^{-4}$ & $p_y$ &$4.04057\times10^{-2}$\\
\hline
$M_h$ & $0.99600$ & $a_h$ & 0.26081 \\
$E_{\rm rad}*1000$ & $4.578\pm 0.684$ & $J_{\rm rad}^z*100$ &
$3.420\pm0.113$\\
$V_{\rm kick}[\KMS]$ & $67\pm35$ & $t_{CAH}$ & $\approx380M$
\end{tabular} \label{table:NID} 
\end{ruledtabular} 
\end{table} 

\begin{figure}
\begin{center}
\includegraphics[width=3.0in]{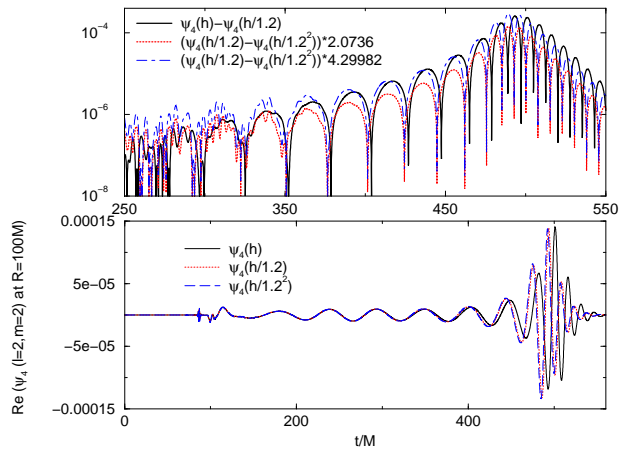}
\caption{Convergence of the real part of the $(\ell=2,m=2)$ mode of $\psi_4$ as
computed by full numerical evolutions and extracted at $R_{\rm obs}=100M$.
 Eighth-order convergence implies $\psi_4(h) -
\psi_4(h/1.2) = 4.29982 (\psi_4(h/1.2) - \psi_4(h/1.2^2))$, while
fourth-order convergence implies
$\psi_4(h) - \psi_4(h/1.2) = 2.0736 (\psi_4(h/1.2) -
\psi_4(h/1.2^2))$.
 Initially, the error in $\psi_4$
is very small and dominated by grid noise, at $t\sim250M$ the
fourth-order convergence of the $\psi_4$ algorithm is apparent, with
the convergence quickly changing to eighth-order as truncation errors
in the metric and extrinsic curvature begin to dominate the errors in
$\psi_4$.  }
\label{fig:NRwaveConv}
\end{center}
\end{figure}

We found that the waveform exhibits eighth-order convergence during
the late-inspiral (when the errors in the waveform are dominated by
phase errors associated with the orbital motion).  The waveform
exhibits noise which tends to mask the convergence at early times. As
the simulation progresses, we first see fourth-order convergence of
the Weyl scalar $\psi_4$ (we used a fourth-order algorithm to compute
$\psi_4$) and later eighth-order convergence, as the
errors in the waveform due to truncation errors in the metric and
extrinsic curvature become important.  In Fig~\ref{fig:NRwaveConv} we
show the convergence of the waveform during the late-inspiral.  Note
the very good agreement of the two finest resolution runs.\\

\noindent
{\it Perturbative evolutions:}
For the case of a nonspinning BHB in
the small mass ratio regime, we can use a Schwarzschild black-hole
(rather than Kerr) as the
background space-time for our perturbation expansion.
 This allows us to
use the simpler metric perturbation equations and the
Regge-Wheeler-Zerilli (RWZ) formalism~\cite{Regge57,Zerilli:1971wd}
\begin{eqnarray}
&& \left[-\frac{\partial^2}{{\partial t}^2} 
+ \frac{\partial^2}{{\partial r^*}^2}-V_\ell^{\rm (even)/(odd)}(r) \right]
\psi^{\rm (even)/(odd)}_{\ell m}(t,r)
\nonumber \\ && 
=S^{\rm (even)/(odd)}_{\ell m}[r_p(t),\Phi_p(t)] \,,
\label{eq:RWZeq}
\end{eqnarray}
where $V_\ell^{\rm (even)/(odd)}$ denotes the Zerilli and
Regge-Wheeler potentials and $S^{\rm (even)/(odd)}_{\ell m}$ is the
source term derived from the particle's trajectory, in the orbital
plane $[r_p(t),\Phi_p(t)]$.  The details are discussed
in~\cite{Lousto:2005ip, Lousto:2005xu}.  and we use the definitions in
\cite{Nakano:2007cj}.  The consistent use of higher order tracks in
this equations has been discussed in~\cite{Mino:2007ft} (See Eq.\
(4.6) therein.)

The coordinates used in numerical simulations are chosen to produce
stable evolutions and  correspond, initially,  to isotropic
coordinates.  Perturbative calculations, on the other hand, regularly
make use of the standard Schwarzschild coordinates. The easiest way to
relate the two is to translate the numerical tracks into the
Schwarzschild coordinates used in Eq.~(\ref{eq:RWZeq}).  This can be
achieved by considering the late-time numerical coordinates that
correspond to radial isotropic `trumpet' stationary $1+\log$ slices of
the Schwarzschild spacetime~\cite{Hannam:2006vv}. We obtain the
explicit coordinate transformations following the procedure detailed
in Ref.~\cite{Brugmann:2009gc}.  We perform the numerical integration
of the perturbative equation (\ref{eq:RWZeq}) with Dirac's delta
sources using the algorithm described in
\cite{Lousto:1997wf,Lousto:2005ip}.

An interesting aspect of the source term in the perturbative evolution
equations (\ref{eq:RWZeq}) is that $S^{\rm (even)/(odd)}_{\ell m}\to0$ 
as $(1-2M/r_p)$
 when the particle
approaches the background horizon (located at $r=2M$). This
leads to essentially free evolution of the Cauchy data once the
particle `sits' on the horizon. In this sense perturbation theory
naturally transitions to a `close limit' evolution. This property
means that once the black holes are close to each other, and form a
common horizon, we no longer need the information of the full
numerical track, since perturbation equations will describe the
radiation essentially in terms of the universal quasi-normal ringing.
In practice, we monitor the numerical tracks and switch off the
source-terms when $r_p(t)\sim 2M$.

\begin{figure}
\begin{center}
\includegraphics[width=3.0in]{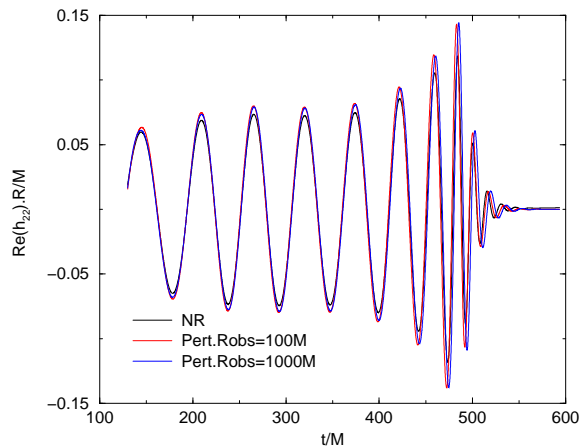}
\caption{The Real part of the $(\ell=2,m=2)$ mode of the strain $h$ as
computed by full numerical evolutions and extracted at $R=100M$, and
the corresponding perturbative evolutions for the same numerical track
extracted at two radii $R_{\rm obs}=100M, 1000M$. The matching of the
numerical and perturbative waveforms ($R_{\rm obs}=100M$ and $1000M$)
for the whole range of evolution for $130.068<t/M<593.564$ are
0.987656 and 0.974751, respectively; while for the two perturbative
waveforms at different extraction radii the matching is 0.938873.  }
\label{fig:NRPC22far}
\end{center}
\end{figure}

\begin{figure}
\begin{center}
\includegraphics[width=3.0in,height=1.2in]{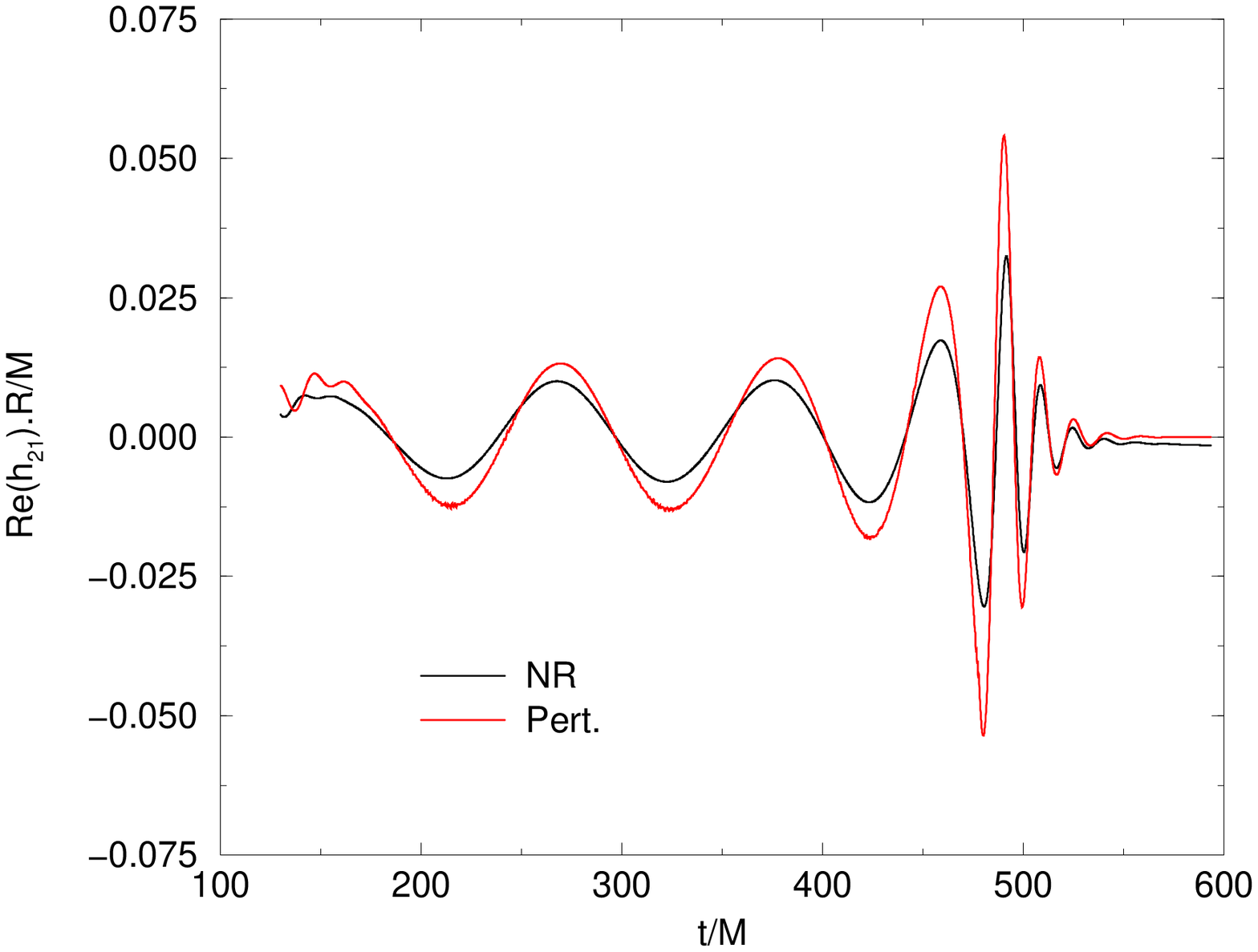}\\
\includegraphics[width=3.0in,height=1.2in]{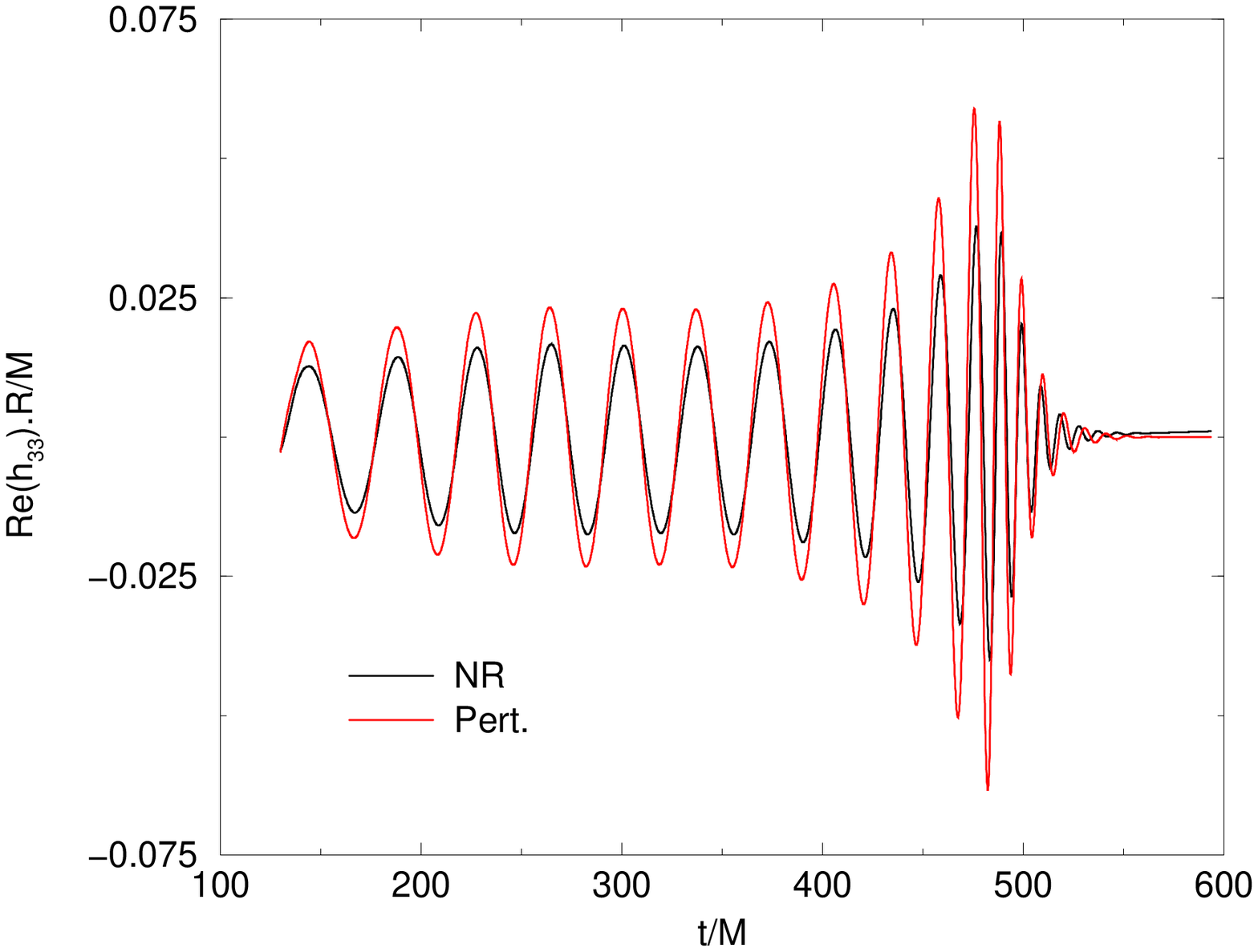}
\caption{Upper: The Real part of the $(\ell=2,m=1)$ mode of the strain
$h$ as computed by a
numerical evolution and extracted at a radius $R_{\rm obs}=100M$, and the
corresponding perturbative (odd parity) evolution for the same numerical 
track extracted at the same radius. 
Lower: The Real part of the $(\ell=3,m=3)$ mode of $h$ as computed by 
the numerical and perturbative evolutions. Similar results are obtained for
the imaginary part of the modes.
}
\label{fig:NRPC2133}
\end{center}
\end{figure}

In Fig.~\ref{fig:NRPC22far}, we plot the leading order $(\ell=2,m=2)$
mode of the strain $h$ calculated both
using the numerical waveform and, perturbatively, from the
numerical trajectories, while in Fig.~\ref{fig:NRPC2133}, 
we compare the nonleading $(\ell=2,m=1)$ and $(\ell=3,m=3)$ modes.
The agreement between waveforms is summarized in Table \ref{tab:match} 
where we computed the overlap between
waveforms as defined in Ref.~\cite{Campanelli:2008nk}.
In addition we evaluate the effects of extraction of the waveform at
finite radii by extracting the perturbative waveform at $R_{\rm
obs}=100M$ and $1000M$ and then (after a time shift)
computing the overlap index (see Table \ref{tab:match}).
The equivalent extraction using standard numerical methods requires
large computational resources, and such extractions have  only
recently been achieved using highly-efficient techniques such as
multi-patch \cite{Pollney:2009ut} and pseudo-spectral
\cite{Scheel:2008rj} methods.

\begin{table} \caption{The overlap (matching) of the real and
imaginary parts  of the modes of the strain $h$. 
The integration time is from $t/M=130.068$ to $593.564$ and 
the definition of the matching is given in Eqs.~(26) and (27) 
of~\cite{Campanelli:2008nk}.}
\label{tab:match}
\begin{ruledtabular}
\begin{tabular}{l|ccc}
   Mode & $\Re (\ell=2, m=2)$ & $\Re (\ell=2, m=1)$ & $\Re (\ell=3, m=3)$ \\
\hline
   Match & 0.987656 & 0.975017 & 0.954138 \\
\hline
\hline
   Mode & $\Im (\ell=2, m=2)$ & $\Im (\ell=2, m=1)$ & $\Im (\ell=3, m=3)$ \\
\hline
   Match & 0.987622 & 0.981982 & 0.953767  \\
  \end{tabular}
  \end{ruledtabular}
\end{table}

\begin{figure}
\begin{center}
\includegraphics[width=3.0in]{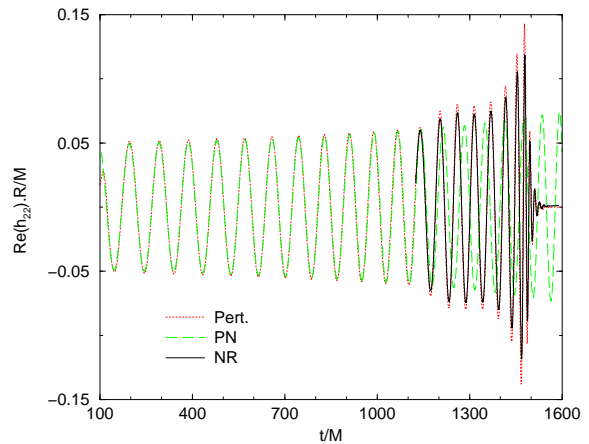}
\caption{The strain computed by a perturbation theory calculation of
using the tracks provided by 3.5PN theory in the early-inspiral, the
numerical tracks during the late-inspiral, and an effective
close-limit calculation post-merger. Note
that we obtain a waveform of duration $\approx1600M$ using only
the information from the first $373M$  of a fully-nonlinear numerical
evolution.
}
\label{fig:PNRNP}
\end{center}
\end{figure}

As we mentioned, the perturbative evolution only makes use of
numerical information (tracks) in the final inspiral and merger phase.
After merger we have free Cauchy evolution of the perturbations. One
can extend this technique to the early inspiral phase, as well. During
the early-inspiral, the binary's
evolution is well described by PN theory and these
trajectories can be used to determine the perturbative waveforms
\cite{Nagar:2006xv}. We can thus, stitch together a 3.5PN
trajectory for the inspiral phase to a full numerical evolution in the
moving puncture approach. In fact, this is how we determined the
initial orbital parameters for the numerical solution of General Relativity
constraints to prepare full numerical evolution. The resulting
waveform is shown in Fig.~\ref{fig:PNRNP}, which also shows
the waveform produced solely with PN methods.
Note that this procedure allows us to  compute the complete waveform
making use of numerical data from the first $t\approx373M$ (of an over $600M$
simulation) of the numerical evolution.
The
procedure resembles the `Lazarus approach' \cite{Baker:2001sf} in
that it makes efficient use of numerical simulations by
restricting them to
the stage of the binary evolution where they are needed.\\
\\
\noindent
{\it Conclusions and Discussion:}
Given the current difficulties in efficiently solving the intermediate 
mass ratio regime by purely nonlinear numerical methods,  we propose
to use perturbation theory to propagate gravitational waves on a black
hole background using trajectory information from a combination of
nonlinear
numerical and semi-analytic methods, such as the post-Newtonian approximation.
In this letter we provided a `proof of principle' comparison between
our hybrid technique and fully nonlinear results in the most challenging case
for perturbation theory, i.e. $q=1/10$.

The benefits of this approach are apparent: i) No fitting parameters
have been used in the generation of the waveforms. ii) Since perturbation
equations naturally transition to close limit ones, i.e. source free
wave equations, numerical tracks are not needed after merger,
thus reducing the cost of the nonlinear simulations
(in our example it represented a $40\%$ reduction in running time.)
iii) The perturbative
evolution also provides a natural way to extract waveforms at very large
radii in a well known and defined gauge, in a region accurately described by
linear perturbations. Thus further cutting the size 
of the required numerical grid and hence the computational cost
of the nonlinear simulation. iv) The computation of even, odd, leading, and
higher-order modes 
$(\ell,m)$ with the same
black holes trajectory. Note that the small amplitude of the
non-leading terms impose
additional requirements of accuracy at the extraction radii and at 
the external boundaries of 
the full numerical grid to avoid interfering reflections.
v) Perturbation codes are highly efficient and run within seconds on
standard
CPUs, and are also amenable for running in GPUs or other accelerated hardware
opening up the possibility of generating online analysis of LIGO/VIRGO data.

In order to improve the accuracy of the generated waveforms we can
consider several extensions to our method. These include perturbations
about a Kerr background using the Teukolsky
equations~\cite{Teukolsky:1973ha}. In addition,
adding second-order perturbative corrections \cite{Campanelli99}
should improve accuracy for larger mass ratios.  Finally, given our
encouraging first results, one can consider a phenomenological
description of intermediate-mass-ratio inspiral trajectories based on
a series of detailed nonlinear  numerical runs to model the tracks of
BHBs as function of mass ratio and spins. The resulting
phenomenological trajectories could then be used to efficiently
generate accurate waveforms over a wide region of the parameter space.

\acknowledgments
We gratefully acknowledge NSF for financial support from grants
PHY-0722315, PHY-0653303, PHY-0714388, PHY-0722703, DMS-0820923, and
PHY-0929114; and NASA for financial support from grants NASA
07-ATFP07-0158 and HST-AR-11763.  Computational resources were
provided by Ranger cluster at TACC (Teragrid allocations TG-PHY080040N
and TG-PHY060027N) and by NewHorizons at RIT.

\bibliographystyle{apsrev}
\bibliography{../../../Bibtex/references}

\end{document}